\documentclass[
 reprint,
 amsmath,amssymb,
 aps,
 pra,
 superscriptaddress,
]{revtex4-2}

\usepackage{float}
\usepackage{setspace}
\usepackage{tabularx}
\usepackage[utf8]{inputenc}
\usepackage{graphicx}
\usepackage{caption}
\usepackage{hyperref}
\usepackage{gensymb}
\usepackage{siunitx}
\usepackage{threeparttable, tablefootnote}
\usepackage{booktabs}
\usepackage{csquotes}
\usepackage[export]{adjustbox}
\usepackage{placeins}
\usepackage{setspace}
\usepackage{subcaption}
\usepackage{gensymb}
\usepackage[usenames,dvipsnames]{color}
\usepackage{cancel}
\usepackage{transparent}
\usepackage{relsize}
\usepackage{graphicx}
\usepackage[percent]{overpic}

\DeclareCaptionJustification{myjust}{\fulljustify}
\makeatletter
\providecommand\fulljustify{%
  \let\\\@centercr
  \leftskip\z@%
  \rightskip\z@%
  \parfillskip\z@\@plus 1fill\relax%
}

\newcommand{\figsize}{1.}
\newcommand{\iso}[2]{\ensuremath{^{#2}\mathrm{#1}}}
\newcommand{\ion}[2]{\ensuremath{^{#2}\mathrm{#1}^+}}
\newcommand{\reportednumber}[1]{\textcolor{black}{#1}}  
\newcommand{\systemnumber}[1]{\textcolor{black}{#1}}  
\newcommand{\invis}[1]{\transparent{0}{#1}}

\newcommand{\CleanoutTransition}[0]{\reportednumber{\SI{373 713.26(2)} {\GHz}}}

\newcommand{\CoolingTransitionNoUnit}[0]{\reportednumber{\SI{640 105.08(2)}{}}}
\newcommand{\ClockTransitionNoUnit}[0]{\reportednumber{\SI{412 017.86(3)} {}}}
\newcommand{\CleanoutTransitionNoUnit}[0]{\reportednumber{\SI{373 713.26(2)} {}}}
\newcommand{\TransitionSevenOhEightNoUnit}[0]{\reportednumber{\SI{423 442.96(3)} {}}}
\newcommand{\RepumpTransitionNoUnit}[0]{\reportednumber{\SI{277 816.92(6)} {}}}
\newcommand{\TransitionEightTwoEightNoUnit}[0]{\reportednumber{\SI{362 288.16(5)} {}}}
\newcommand{\TransitionThreeEightTwoNoUnit}[0]{\reportednumber{\SI{785 731.12(4)} {}}}
\begin{document}

\preprint{AIP/123-QED}

\title{Spectroscopy of electric dipole and quadrupole transitions in \ion{Ra}{224}}

\renewcommand*{\thefootnote}{\relax}
\newcommand{\UCSBAffiliation}[0]{Department of Physics, University of California, Santa Barbara, California 93106, USA}
\newcommand{\UCSBAffiliationCS}[0]{Department of Computer Science, University of California, Santa Barbara, California 93106, USA}

\author{Spencer Kofford}
\email{sjkofford@ucsb.edu}
\affiliation{\UCSBAffiliation}
\author{Haoran Li}
\affiliation{\UCSBAffiliation}
\author{Robert Kwapisz}
\affiliation{\UCSBAffiliation}
\author{Roy A. Ready}
\affiliation{\UCSBAffiliation}
\author{Akshay Sawhney}
\affiliation{\UCSBAffiliation}
\author{Oi Chee Cheung}
\affiliation{\UCSBAffiliationCS}
\author{Mingyu Fan}
\affiliation{\UCSBAffiliation}
\author{Andrew M. Jayich}
\affiliation{\UCSBAffiliation}

\date{\today}

\begin{abstract}
We report on spectroscopy of the low-lying electronic transitions in \iso{Ra}{224}$^+$. We measured the frequencies of the $^2{S}_{1/2} \ $$\leftrightarrow$$\  ^2{P}_{1/2}$ cooling transition, the $^2{S}_{1/2}\ $$\leftrightarrow$$\ ^2{D}_{5/2}$ clock transition, the $^2{D}_{3/2} \ $$\leftrightarrow$$\  ^2{P}_{3/2}$ electric dipole transition, and the $^2{D}_{5/2} \ $$\leftrightarrow$$\  ^2{P}_{3/2}$ cleanout transition. From these measurements we calculate the frequencies of the $^2{D}_{3/2}\ $$\leftrightarrow$$\ ^2{P}_{1/2}$ repump transition, the $^2{S}_{1/2} \ $$\leftrightarrow$$\  ^2{D}_{3/2}$ electric quadrupole transition, and the $^2{S}_{1/2} \ $$\leftrightarrow$$\  ^2{P}_{3/2}$ electric dipole transition. The ion's low charge-to-mass ratio and convenient wavelengths make \ion{Ra}{224} a promising optical clock candidate.
\end{abstract}
\maketitle

\section{Introduction}
\begin{figure}
    \centering
    \captionsetup{justification=myjust,singlelinecheck=false}
    \includegraphics[width=\figsize\linewidth]{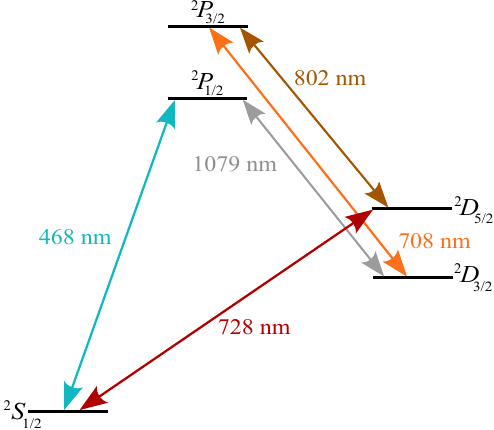}
    \caption{Laser wavelengths used to drive \ion{Ra}{224} transitions in this work. The 708-nm laser is not typically used for ion control, but measuring the $^2D_{3/2}\ $$\leftrightarrow$$ \ ^2P_{3/2}$ transition frequency allows us to calculate the $^2D_{3/2}\ $$\leftrightarrow$$ \ ^2P_{1/2}$ repump transition frequency. }
    \label{fig:Ra_level_structure}
\end{figure}

The radium ion is an alkali-like atom that can be controlled with accessible laser wavelengths (see Fig. \ref{fig:Ra_level_structure}).  The \mbox{radium-224} isotope (\reportednumber{3.6} d half-life) has a simple level structure and no nuclear spin ($I=0$). Despite radium-224's short half-life, we have been working with this isotope in a sealed vacuum system for over a year.  This is enabled by a radium oven which is continuously populated by the decay of \mbox{thorium-228} (\reportednumber{1.9} y half-life) to \mbox{radium-224} that when heated generates a radium beam for photoionization and trapping \cite{Fan2023a}.  The lack of nuclear spin makes optical control easier compared to isotopes with nuclear spin, such as radium-225 ($I=1/2$).  Though radium-224 has a much shorter half-life than radium-226 (1,600 y),  it has the advantage that the background pressure due to its radon-220 daughter ($\tau=55.6$ s) can be much less than the radium-226 daughter, radon-222 ($\tau=3.8$ d). \ion{Ra}{224} is a logic ion option for quantum logic spectroscopy of heavy spectroscopy ions \cite{Schmidt2005, Chou2017}, such as molecules for testing time-reversal symmetry \cite{Yu2021, Fan2021}.  In addition to convenient wavelengths, the radium ion's high mass and low charge also make it an appealing optical clock candidate by reducing uncertainty due to the second-order Doppler effect \cite{Berkeland1998, Holliman2022}.   The thermal contribution to the second-order Doppler shift decreases linearly in mass.  If there is uncompensated micromotion, for example due to trap imperfections, the second-order Doppler shift drops quadratically with the ion's charge-to-mass ratio.

\begin{table}
    \caption{\label{table:transitions} Summary of \ion{Ra}{224} transition frequencies. Asterisks indicate frequencies calculated from the measured transitions.
    }
    \begin{tabularx}{\columnwidth}{X >{\centering\arraybackslash}X >{\centering\arraybackslash}X}
        \toprule
         {Transition} & Wavelength [nm] & Frequency [GHz] \\
        \midrule
        $^2{S}_{1/2}~$$\leftrightarrow$$~^2{P}_{3/2}$ & 382 & \TransitionThreeEightTwoNoUnit \text{*} \\
        $^2{S}_{1/2}~$$\leftrightarrow$$~^2{P}_{1/2}$ & 468 & \CoolingTransitionNoUnit \invis{*}\\
        $^2{D}_{3/2}\hspace{0.4mm}$$\leftrightarrow$$~^2{P}_{3/2}$ & 708 & \TransitionSevenOhEightNoUnit\invis{*} \\        
        $^2{S}_{1/2}~$$\leftrightarrow$$~^2{D}_{5/2}$ & 728 & \ClockTransitionNoUnit \invis{*}\\
        $^2{D}_{5/2}\hspace{0.5mm}$$\leftrightarrow$$~^2{P}_{3/2}$ & 802 & \CleanoutTransitionNoUnit \invis{*}\\
        $^2{S}_{1/2}~$$\leftrightarrow$$~^2{D}_{3/2}$ & 828 & \TransitionEightTwoEightNoUnit \text{*}\\
        $^2{D}_{3/2}\hspace{0.5mm}$$\leftrightarrow $$~^2{P}_{1/2}$ & 1079 & \RepumpTransitionNoUnit \text{*}\\
        \bottomrule
    \end{tabularx}
\end{table}

In this paper we report on measurements of the \ion{Ra}{224} transition frequencies for laser cooling, the $^2S_{1/2}\ $$\leftrightarrow$$\ ^2D_{5/2}$ narrow electric quadrupole transition, the $^2D_{3/2}\ $$\leftrightarrow$$\ ^2P_{3/2}$ transition, and the $^2D_{5/2}\ $$\leftrightarrow$$\ ^2P_{3/2}$ cleanout transition. Tellurium (Te$_2$) and iodine (I$_2$) molecular vapor cells are used as absolute frequency references for the radium transitions. The transition frequencies are obtained by recording the frequency difference between the radium transition and a molecular absorption peak. We then determine the radium transition frequency from the value of the molecular reference given in IodineSpec5 \cite{Knöckel2004} or the tellurium atlas \cite{Cariou1980}.

\begin{figure}[t]
    \centering
    \includegraphics[width=\figsize\linewidth]{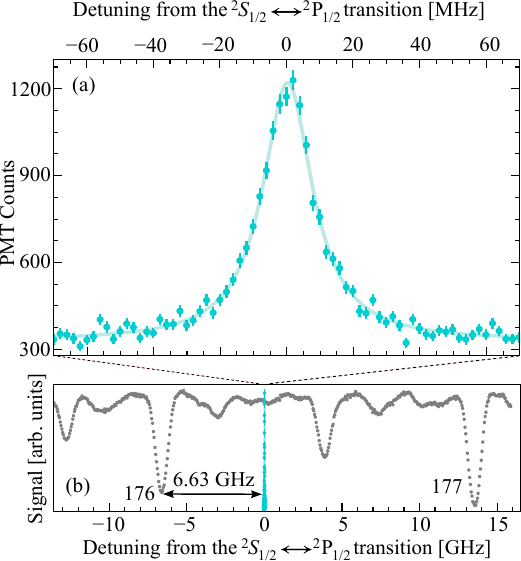}
    \captionsetup{justification=myjust,singlelinecheck=false}
    \caption{(a) Spectroscopy of the $^2{S}_{1/2} \ $$\leftrightarrow$$\  ^2{P}_{1/2}$ transition. The Lorentzian fit gives a full width at half maximum (FWHM) of \reportednumber{18.3(5)} MHz. (b) The tellurium absorption peaks 176 and 177 (gray) with $^2{S}_{1/2} \ $$\leftrightarrow$$\  ^2{P}_{1/2}$ spectroscopy data overlaid.}
    \label{fig:468_combined_2}
\end{figure}

\begin{figure}
    \centering
    \vspace{+2mm}
    \includegraphics[width=\figsize\linewidth]{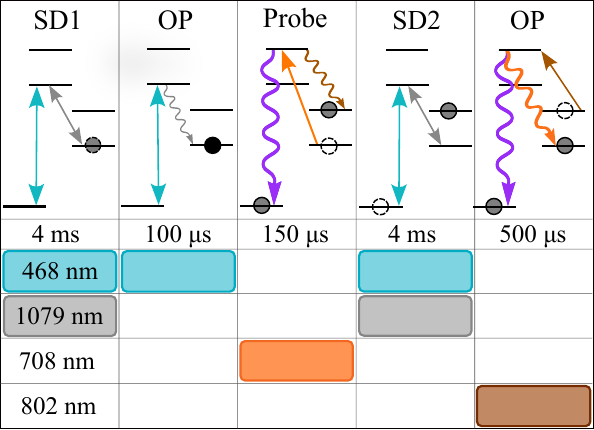}
    \captionsetup{justification=myjust,singlelinecheck=false}
    \caption{$^2{D}_{3/2} \ $$\leftrightarrow$$\  ^2{P}_{3/2}$ spectroscopy pulse sequence. }
    \label{fig:708-sequence}
\end{figure}

With these four measurements we are able to determine the frequencies of the five electric dipole and two electric quadrupole transitions between the lowest-energy states of $^{224}$Ra$^+$ (see Table \ref{table:transitions}).

\section{Experimental Setup}

\begin{figure}[b]
    \centering
    \includegraphics[width=\figsize\linewidth]{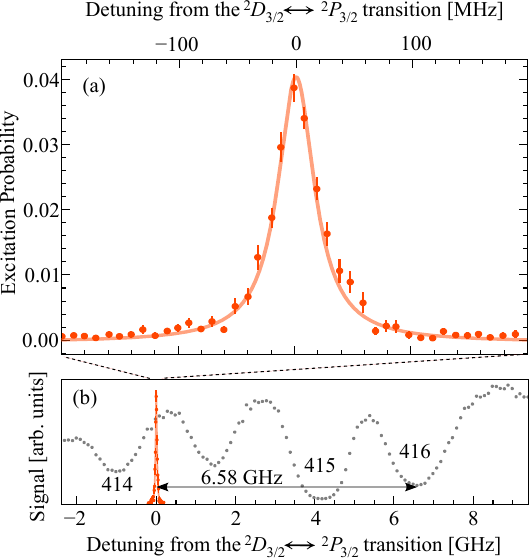}
    \captionsetup{justification=myjust,singlelinecheck=false}
    \caption{(a) Spectroscopy of the $^2{D}_{3/2} \ $$\leftrightarrow$$\  ^2{P}_{3/2}$ transition. The excitation probability is plotted as a function of 708-nm probe laser detuning. A Lorentzian fit gives a FWHM of \reportednumber{38.3(1.8)} MHz. (b) The $^2{D}_{3/2}\ $$ \leftrightarrow$$\ ^2{P}_{3/2}$ radium transition (red) is plotted with nearby iodine absorption peaks 414, 415, and 416 \cite{Gerstenkorn1982a}.}
    \label{fig:708_combined_2}
\end{figure}

The measurements use a single ion in a linear Paul trap with an rf electrode separation of \reportednumber{6} mm and an axial separation of \reportednumber{15} mm, described in Fan \textit{et al.} \cite{Fan2019}.  Ions are loaded by first driving the 483-nm $^1{S}_{0}\ $$\leftrightarrow$$\ {^1P_0}$ transition of radium atoms in an atomic beam and then photoionizing atoms in the $^1P_0$ state with non-resonant light at 450-nm \cite{Fan2023a}. 

The ion is Doppler cooled with 468-nm light red-detuned from resonance and repumped with 1079-nm light. The 468- and 1079-nm lasers are locked to optical cavities, and the 708-, 728-, and 802-nm lasers are locked to a wavemeter (High Finesse WS-8, 10 MHz uncertainty). The frequency and intensity of laser light that addresses the radium ion is controlled with double-passed acousto-optic modulators (AOMs). 

For each transition measurement the probe laser light passes through a molecular reference cell where spectroscopic lines are used as absolute frequency references.  We use a tellurium absorption line to anchor the $^2S_{1/2}\ $$\leftrightarrow$$\ ^2P_{1/2}$ line center.  For the transitions in the red (708-, 728-, and 802-nm) we use iodine as the frequency reference.  The molecular spectroscopy light is split into two beams.  The first beam passes through the vapor cell where absorption peaks are recorded with a photodiode and fit with Voigt functions.  The second beam goes to another photodiode to record the laser's intensity, which is used to calibrate the molecular absorption peak signal.  The molecular spectroscopy is done before and after the $^{224}$Ra$^+$ spectroscopy in order to reduce the effect of wavemeter drift.  The reference peak frequency is taken as the averaged value of the two measurements.  The scanned frequency range is chosen to include at least two peaks to identify the molecular reference lines. 

We ascribe a total uncertainty to each radium transition of \mbox{$\sigma_{\mathrm{tot}} = \sqrt{ \sigma_{\mathrm{wm}}^2 + \sigma_{\mathrm{ref}}^2 + \sigma_{\mathrm{Ra}}^2 + \sigma_{\mathrm{mol}}^2} $}.  The wavemeter uncertainty $\sigma_{\mathrm{wm}}$ is \systemnumber{10} MHz for all measurements. Uncertainties from fitting the radium transitions and molecular absorption peaks are $\sigma_{\mathrm{Ra}}$ and $\sigma_{\mathrm{mol}}$, and the molecular spectroscopy reference uncertainty is $\sigma_{\mathrm{ref}}$. For each molecular reference line we check magnetic field dependence by varying the magnetic field from -\systemnumber{4} G to +\systemnumber{4} G and do not measure any change in the line's center frequency.  Similarly, we vary the temperature of the iodine and tellurium cells by $\pm$\systemnumber{50} \textdegree C and $\pm$\systemnumber{100} \textdegree C, and find no change in the center frequencies of all reference peaks.

\section{Radium Spectroscopy}

\begin{figure}
    \centering
    \vspace{+8.5mm}
    \includegraphics[width=\figsize\linewidth]{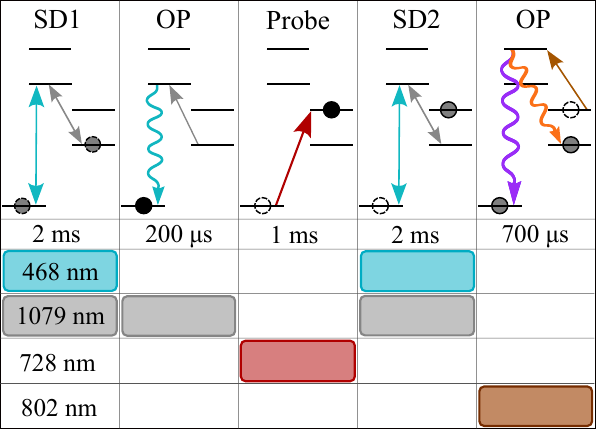}
    \captionsetup{justification=myjust,singlelinecheck=false}
    \caption{$^2{S}_{1/2} \ $$\leftrightarrow$$\  ^2{D}_{5/2}$ spectroscopy pulse sequence. } 
    \label{fig:728-sequence}
\end{figure}

The $^2{S}_{1/2} \ $$\leftrightarrow$$\  ^2{P}_{1/2}$ and $^2{D}_{3/2} \ $$\leftrightarrow$$\  ^2{P}_{1/2}$ transitions of \ion{Ra}{224} are used to Doppler cool the ion. If population driven into the $^2{P}_{1/2}$ state decays to the ground state, a 468-nm photon is emitted.  A fraction of these photons are collected by our photomultiplier tube (PMT). 

Ions are initialized in the $^2 S_{1/2}$ and $^2D_{3/2}$ states which is confirmed at the start of each spectroscopy pulse sequence with state detection (SD).  The SD consists of the cooling (468-nm) and repump (1079-nm) lasers, which cause the ion to fluoresce if population is in the $^2{S}_{1/2}$ or $^2{D}_{3/2}$ state.  We refer to these two states as bright states and the $^2D_{5/2}$ state as the dark state as the ion does not fluoresce if it is in this state. For each millisecond of state detection we collect \reportednumber{22} photons on average if the ion is in a bright state and \reportednumber{two} photons if population is in the dark state.  If the first SD (SD1) measures low PMT counts, the corresponding data point is discarded because the ion was not properly initialized.  SD1 is followed by an optical pumping (OP) step that prepares population into the desired state.  A final state detection (SD2) discriminates if the probe pulse drove the transition. All pulse sequences end with an optical pumping step which returns population from the $^2{D}_{5/2}$ dark state back to the cooling cycle. 

\begin{figure}
    \centering
    \includegraphics[width=\figsize\linewidth]{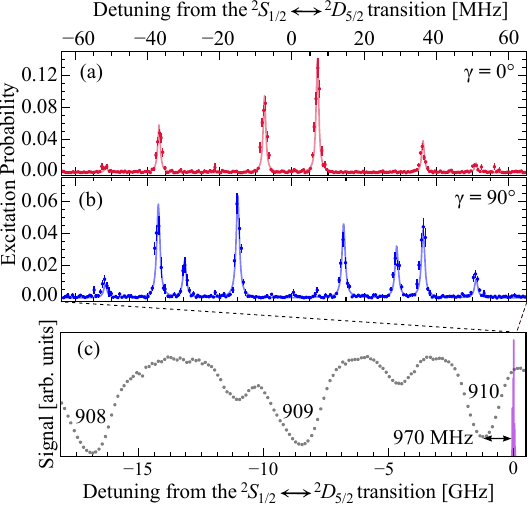}
    \captionsetup{justification=myjust,singlelinecheck=false}
    \caption{The ten $^2{S}_{1/2}\ $$\leftrightarrow $$\ ^2{D}_{5/2}$ Zeeman transitions. (a) Spectroscopy with $\gamma$ = \systemnumber{0\textdegree} where $\Delta m = \pm 1$ transitions are suppressed. (b) Spectroscopy with $\gamma$ = \systemnumber{90\textdegree} where $\Delta m = 0$ transitions are suppressed. The reported $^2{S}_{1/2}\ $$\leftrightarrow $$\ ^2{D}_{5/2}$ frequency is the weighted sum of the $\gamma$ = \systemnumber{0\textdegree} and $\gamma$~=~\systemnumber{90\textdegree} scan center frequencies. (c) The combined spectroscopy of (a) and (b) (purple) plotted with the $\mathrm{I}_{2}$ reference spectroscopy (gray). }
    \label{fig:728_combined_2}
\end{figure}

\section{$^2{S}_{1/2} \ $$\leftrightarrow$$\  ^2{P}_{1/2}$ Spectroscopy}

To measure the $^2{S}_{1/2} \ $$\leftrightarrow$$\  ^2{P}_{1/2}$ transition at 468-nm, a two step pulse sequence is repeated while the laser is scanned across the resonance with random frequency ordering (see Fig.~\ref{fig:468_combined_2}). The sequence consists of a \systemnumber{\SI{3}{\micro\second}} pulse of 468-nm light followed by a \systemnumber{\SI{3}{\micro\second}} pulse of 1079-nm light.  A transition linewidth of \reportednumber{18.3(5)} MHz is recorded for this transition, which sets a new lower bound on the $ ^2{P}_{1/2}$ state lifetime of \reportednumber{8.7(2)} ns, which is consistent with the calculated value of 8.72 ns  \cite{Pal2009}. 

The cooling transition is referenced to tellurium absorption line~176 \cite{Cariou1980} which has an uncertainty of $\sigma_{\mathrm{ref}} = $ \systemnumber{23} MHz. The fit uncertainties are $\sigma_{\mathrm{mol}} =$ \reportednumber{1.06} MHz and $\sigma_{\mathrm{Ra}} =$ 
 \reportednumber{150} kHz.

\section{$^2{D}_{3/2} \ $$\leftrightarrow$$\  ^2{P}_{3/2}$ Spectroscopy}

\begin{figure}
    \centering
    \vspace{+8.5mm}
    \includegraphics[width=\figsize\linewidth]{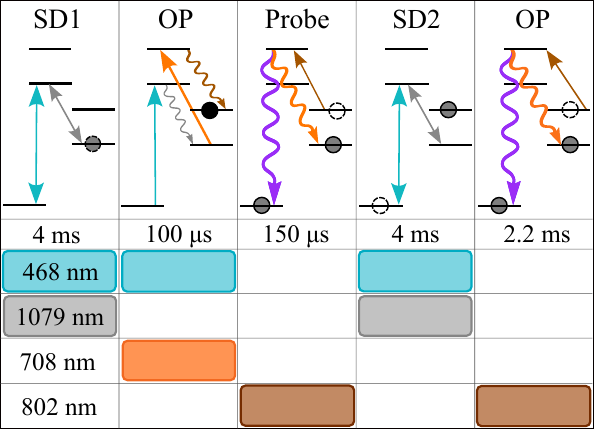}
    \captionsetup{justification=myjust,singlelinecheck=false}
    \caption{$^2{D}_{5/2} \ $$\leftrightarrow$$\ ^2{P}_{3/2}$ spectroscopy pulse sequence. }
    \label{fig:802-sequence}
\end{figure}

The pulse sequence for $^2{D}_{3/2}\ $$ \leftrightarrow$$\ ^2{P}_{3/2}$ spectroscopy, shown in Fig.~\ref{fig:708-sequence}, is repeated while the 708-nm probe laser is manually scanned over the transition (see Fig. ~\ref{fig:708_combined_2}). The $^2{D}_{3/2}\ $$\leftrightarrow$$\ ^2{P}_{3/2}$ transition is referenced to iodine line~416 ~\cite{Gerstenkorn1982a} which has an uncertainty of $\sigma_{\mathrm{ref}} = $ \reportednumber{30} MHz from IodineSpec5. The fit uncertainties are $\sigma_{\mathrm{mol}} =$ \reportednumber{2.85} MHz and $\sigma_{\mathrm{Ra}} =$ \reportednumber{720} kHz.

\section{$^2{S}_{1/2} \ $$\leftrightarrow$$\  ^2{D}_{5/2}$ Spectroscopy}

The $^2{S}_{1/2}\ $$\leftrightarrow $$\ ^2{D}_{5/2}$ spectroscopy pulse sequence shown in Fig.~\ref{fig:728-sequence} is repeated while the 728-nm probe laser is manually swept over the resonance, driving all ten transitions between Zeeman levels of the $^2{S}_{1/2} $ and $^2{D}_{5/2}$ states.

To resolve the Zeeman transitions shown in Fig.~\ref{fig:728_combined_2}, a magnetic field is applied at $\phi =$ \systemnumber{45\textdegree} with respect to the 728-nm laser's $k$ vector. The 728-nm laser polarization forms an angle $\gamma$ with respect to the magnetic field vector projected into the plane of incidence. Two values of $\gamma$ are used to drive all ten Zeeman transitions. In the first measurement $\gamma$ is set to \systemnumber{0\textdegree}, where $\Delta m = \pm$ 1 transitions are suppressed. In the second measurement $\gamma$ is set to \systemnumber{90\textdegree}, and $\Delta m$ = 0 transitions are suppressed.

The clock transition is referenced to iodine line~910 \cite{Gerstenkorn1982a} which has an uncertainty of $\sigma_{\mathrm{ref}} = $ \reportednumber{30} MHz from IodineSpec5. The fit uncertainties are $\sigma_{\mathrm{mol}} =$ \reportednumber{1.33} MHz and $\sigma_{\mathrm{Ra}} =$ \reportednumber{10} kHz.

\section{$^2{D}_{5/2} \ $$\leftrightarrow$$\  ^2{P}_{3/2}$ Spectroscopy}

\begin{figure}
    \centering
    \includegraphics[width=\figsize\linewidth]{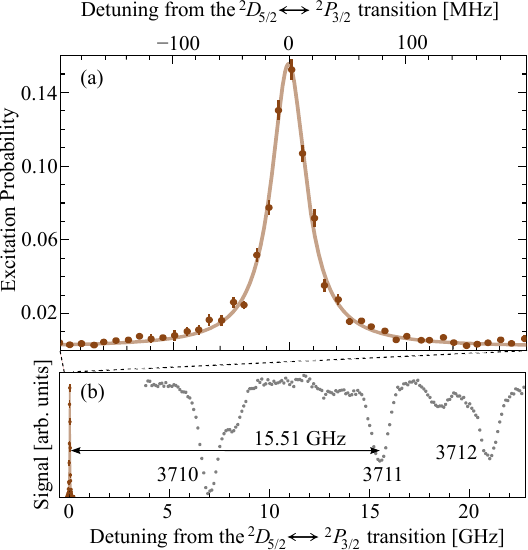}
    \captionsetup{justification=myjust,singlelinecheck=false}
    \caption{(a) Spectroscopy of the $^2{D}_{5/2} \ $$\leftrightarrow$$\  ^2{P}_{3/2}$ transition. The excitation probability is plotted as a function of the 802-nm laser detuning. A Lorentzian fit gives a FWHM of \reportednumber{36.0(1.0)} MHz. (b) The $^2{D}_{5/2}\ $$ \leftrightarrow$$\ ^2{P}_{3/2}$ radium transition (brown) is plotted with nearby iodine absorption peaks \cite{Gerstenkorn1982a}. }
    \label{fig:802_combined_2}
\end{figure}

The $^2{D}_{5/2}\ $$\leftrightarrow$$\ ^2{P}_{3/2}$ pulse sequence, depicted in Fig.~\ref{fig:802-sequence}, is repeated while the 802-nm laser's frequency is manually tuned over the resonance (see Fig.~\ref{fig:802_combined_2}). We report the $^2{D}_{5/2} \ $$\leftrightarrow$$\  ^2{P}_{3/2}$ transition center frequency to be \CleanoutTransition. The $^2{D}_{5/2} \ $$\leftrightarrow$$\  ^2{P}_{3/2}$ cleanout transition is referenced to iodine peak 3711 in ref.~\cite{Gerstenkorn1982a} which has an uncertainty of \reportednumber{2} MHz from IodineSpec5. Fitting uncertainties for the iodine and radium peaks are \reportednumber{4.29} MHz and \reportednumber{400} kHz.

\section{Calculated Frequencies}

The $^2{D}_{3/2} \ $$\leftrightarrow$$\  ^2{P}_{1/2}$ transition repumps population to the ground state during laser cooling.  The $^2{D}_{3/2} \ $$\leftrightarrow$$\  ^2{P}_{1/2}$ frequency is not close to any peaks in the iodine absorption spectrum, but with our measurements we can calculate the transition frequency \mbox{$\nu_{1079} = \nu_{468} - ( \nu_{728} + \nu_{802} -\nu_{708} ) $}. The uncertainty is the quadrature sum of the measured transition uncertainties, $ \sigma_{\mathrm{1079}} = \sqrt{ \sigma_{\mathrm{468}}^2 + \sigma_{\mathrm{708}}^2 + \sigma_{\mathrm{728}}^2 + \sigma_{\mathrm{802}}^2 } $. The $^2{S}_{1/2}\ $$\leftrightarrow$$ \ ^2{P}_{3/2}$ and $^2{S}_{1/2} \ $$\leftrightarrow $$\ ^2{D}_{3/2}$ transition frequencies are determined in a similar fashion.

\section{Conclusion}

We report frequencies for the low-lying optical transitions in \ion{Ra}{224} that are useful for controlling the ion.  Accounting for isotope shifts, these measurements are consistent with frequency measurements of \ion{Ra}{226} ~\cite{Fan2019, Holliman2019}.  The reported \ion{Ra}{224} frequency uncertainties are all below 100~MHz, which is sufficient to start working with radium-224 ions.  These frequencies along with the previous measurements and the long-lived radium source ~\cite{Fan2023a} lays the foundation for using \ion{Ra}{224} in optical clocks, quantum information science, and quantum logic spectroscopy.

\section*{Acknowledgements}

We thank Samuel Brewer for helpful discussions.  H.L. was supported by ONR Grant No.~N00014-21-1-2597 and M.F. was supported by DOE Award No.~DE-SC0022034. S.K., R.K., R.A.R., A.S., and A.M.J. were supported by the Heising-Simons Foundation Award No.~2022-4066, the W.M. Keck Foundation, NIST Award No.~60NANB21D185, NSF NRT Award No.~2152201, the Eddleman Center, the Noyce Initiative, and NSF Award Nos.~2326810, and N2146555.  The isotope used in this research was supplied by the U.S. Department of Energy Isotope Program, managed by the Office of Isotope R\&D and Production.

\end{document}